\documentclass{appolb}
\usepackage{epsfig}
\usepackage{amssymb}
\usepackage{color}

\begin{document}

\title{Current log-periodic view on future world market development}

\author{Stanis{\l}aw Dro\.zd\.z$^{1,2}$, Jaros{\l}aw Kwapie\'n$^{1}$, 
Pawe{\l} O\'swi\c ecimka$^{1}$, Josef Speth$^{1,3}$
\address{$^1$Institute of Nuclear Physics, Polish Academy of Sciences,
PL--31-342 Krak\'ow, Poland \\
$^2$Institute of Physics, University of Rzesz\'ow, PL--35-310 Rzesz\'ow,
Poland \\
$^3$Institut f\"ur Kernphysik, Forschungszentrum J\"ulich, D--52425 J\"ulich,
Germany}}

\maketitle

\begin{abstract}
Applicability of the concept of financial log-periodicity is discussed
and encouragingly verified for various phases of the world stock markets development 
in the period 2000-2010. In particular, a speculative forecasting scenario designed 
in the end of 2004, that properly predicted the world stock market increases in 2007,
is updated by setting some more precise constraints on the time of duration of the
present long-term equity market bullish phase.
A termination of this phase is evaluated to occur in around November 2009. 
In particular, on the way towards this dead-line, a Spring-Summer 2008 increase
is expected. On the precious metals market a forthcoming critical time signal
is detected at the turn of March/April 2008 which marks a tendency for 
at least a serious correction to begin. 

In the present extended version some predictions 
for the future oil price are incorporated. In particular a serious correction 
on this market is expected in the coming days.
\end{abstract}



\vskip1.0cm

{\it E-mail address: Stanislaw.Drozdz@ifj.edu.pl}

\section{Introduction}

The financial dynamics is a multiscale phenomenon and therefore
the question which of its properties are scale invariant and which
are scale characteristic refers to the essence of this phenomenon.
There exists strong related evidence that at least a large portion
of the financial dynamics is governed by phenomena analogous to
criticality in the statistical physics sense~\cite{sornette_cc}.
In its conventional form criticality implies a scale invariance~\cite{stanley} 
of a properly defined function $\Phi(x)$ characterizing the system:  
\begin{equation}
\Phi(\lambda x) = \gamma \Phi(x).
\label{eq:F}
\end{equation}
A constant $\gamma$ then describes how the properties of the system 
change under rescaling by the factor $\lambda$.
The general solution to this equation reads~\cite{nauenberg}:
\begin{equation}
\Phi(x) = x^{\alpha} \Pi(\ln(x)/\ln(\lambda)),
\label{eq:FP}
\end{equation}
where the first term represents a standard power-law that is characteristic 
of continuous scale-invariance with the critical exponent 
$\alpha = \ln(\gamma) / \ln(\lambda)$ and $\Pi$ denotes a periodic function
of period one. This general solution can be interpreted in terms of
discrete scale invariance. Due to the second term the conventional 
dominating scaling acquires a correction that is periodic in $\ln(x)$.
Imprints of such oscillations can often be identified in the financial 
dynamics~\cite{sornette_jp,feigenbaum,vandevalle,drozdz99}. 
To make a proper mapping one defines $x = \vert T - T_c \vert$, where $T$ denotes
the clock time labelling the original price time series and represents 
a distance to the critical point $T_c$. 
The emerging sequence of spacings between the
corresponding consecutive repeatable structures at $x_n$ - as seen 
in the linear scale - forms a geometric progression according to the relation
$(x_{n+1}-x_n) / (x_{n+2}-x_{n+1}) = \lambda$.
The time points $T_c$ thus correspond to the accumulation of such oscillations 
and, in the context of the financial dynamics such points mark a reversal 
of the trend.
An important related element, for 
a proper interpretation and handling of the financial patterns as well as for 
consistency of the theory, is that such log-periodic oscillations manifest
their action self-similarly through various time scales~\cite{drozdz99}. 
This applies both to the log-periodically 
accelerating bubble market phase as well as to the 
log-periodically decelerating anti-bubble phase.

%-------------------------------Figure 1-----------------------------------------
\begin{figure}[h]
\begin{center}
\includegraphics[width=0.7\textwidth,height=0.6\textwidth]{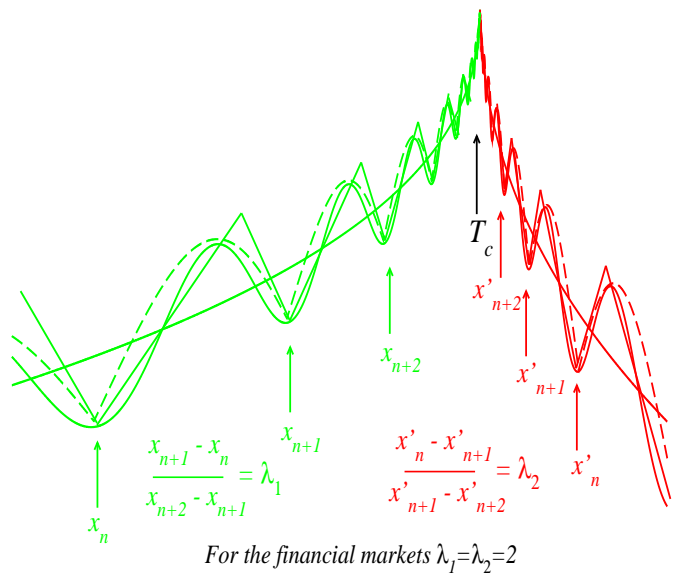}
\caption{Schematic illustration of the possible structures, relevant
for the financial markets and consistent with the Eq.~\ref{eq:FP}.
The thick - cusp-shaped at $T_c$ - solid line corresponds to the conventional
phase transition and in the financial context it reflects an overall market trend.
Superimposed on top of this trend are three - cosine, cosine modulus and saw-like -
possible practical representations for the oscillation pattern periodic in the 
logarithm of the distance $x$ from the critical time $T_c$ 
($x = \vert T - T_c \vert$). Log-periodicity means that 
the ratios $\lambda_1$ and $\lambda_2$ of the distances between the consecutive 
repeatable structures $x_n$ are constant. Furthermore, for the financial markets
$\lambda_1 = \lambda_2 = 2$.}       
\end{center}
\end{figure}
%----------------------------------------------------------------------------------

Fig.~1 schematically illustrates the relevant structures on one 
particular time scale. The thick solid line
corresponds to the first term $(x^{\alpha})$ in Eq.~\ref{eq:FP} and it represents
the global market trends on both sides of $T_c$, 
increasing and decreasing respectively. On both these sides the log-periodic oscillations
are superimposed, accelerating and decelerating correspondingly.
These oscillations are generated by the second term in Eq.~\ref{eq:FP}.
A specific form of the periodic function $\Pi$ in this Equation is as yet
not provided by any first principles.
Since in the corresponding methodology the oscillation structure carries
the most relevant information about the market dynamics, for transparency 
one uses the simplest representations for such a function. One reasonable
possibility is the first term of its Fourier expansion,  
\begin{equation}
\Pi(\ln(x)/\ln(\lambda)) = A + B \cos({\omega \over 2\pi} \ln(x) + \phi). 
\label{eq:FPE}
\end{equation}
This implies that $\omega = 2\pi / \ln(\lambda)$. Already such a simple 
parametrization allows to properly reflect the contraction of oscillations,
especially on the larger time scales. On the smaller time scales just 
replacing the {\it cosine} by its modulus equally well describes oscillations
and in addition it often even better follows departures of the market amplitude 
from its average trend. Another simple - from the market perspective
an even more realistic - representation of $\Pi$ is to use an asymmetric 
saw-like function. Such three possibilities are indicated in Fig.~1.
What they all have however to obey is the same contraction ratio (preferred
scaling factor) expressed by $\lambda_1$ and $\lambda_2$. For the real markets
more and more evidence is collected that the preferred scaling factor $\lambda \approx 2$
and is common to all the scales and markets~\cite{drozdz03} both, in the
log-periodically accelerating bubble as well as in the log-periodically decelerating
anti-bubble phases~\cite{bartolozzi}. We thus set $\lambda_1 = \lambda_2 = \lambda = 2$. 
Universality of the $\lambda$, establishes very valuable
and in fact crucial constraint on possible forms of the analytic 
representations of the market trends and of the oscillation patterns, 
including the future ones. This greatly amplifies a predictive power
of the corresponding methodology.
Also very helpful in this respect is the requirement of self-similarity which
helps to clarify the significance of a given pattern and allows to determine
on what time scale it operates.  
The present contribution provides further arguments in this favour.

\section{World stock market since 2003}

As far as the trends are concerned the contemporary stock market 
indices world-wide are remarkably synchronized. This is one of the 
characteristics reflecting the world globalization~\cite{drozdz01}. 
For the recession period since 2000 this effect is illustrated 
in ref.~\cite{bartolozzi}. This period ends approximately in the first 
months of 2003 and the stock markets synchronously enter the 
bullish period. As is shown in Fig.~2 all the indices assume the up trend. 
There of course are some differences in the magnitude of oscillation 
amplitudes but the common similarity of this oscillation pattern is 
clearly visible. Due to this similarity in the following we shall concentrate
mostly on the American S$\&$P500 index because it represents the world 
largest market and is thus expected to provide the most reliable indicator 
of the global world market trends.

%--------------------------------Figure 2--------------------------------------
\begin{figure}
\begin{center}
%\vspace{1cm} 
%\epsffile{fig1.eps}
%\centerline{\epsfig{figure=fig2.eps,height=8cm,width=10cm}}
\includegraphics[width=0.7\textwidth,height=0.6\textwidth]{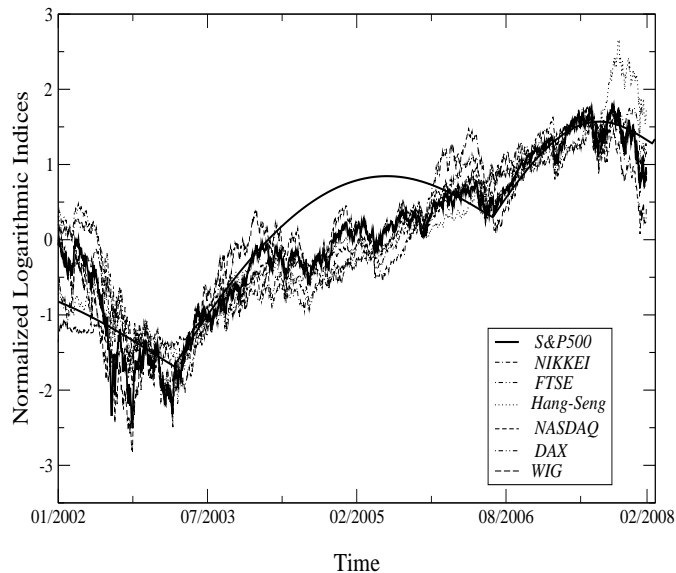}
\caption{ Time series of the logarithm of the most important indices world-wide 
in the period 2002-2007. The time series have been appropriately normalized, 
$P(t) \rightarrow \frac{P(t)-<P(t)>}{\sigma(P(t))}$, 
where $<...>$ denotes the average
over the period under consideration and $\sigma$ is the standard deviation.}
\label{fig2}
\end{center}
\end{figure}
%------------------------------------------------------------------------------

%--------------------------------Figure 3--------------------------------------
\begin{figure}[th]
%\vspace{1cm} 
%\centerline{\epsfig{figure=fig3.eps,height=8cm,width=10cm}}
%\epsffile{fig1.eps}
\includegraphics[width=0.5\textwidth,height=0.5\textwidth]{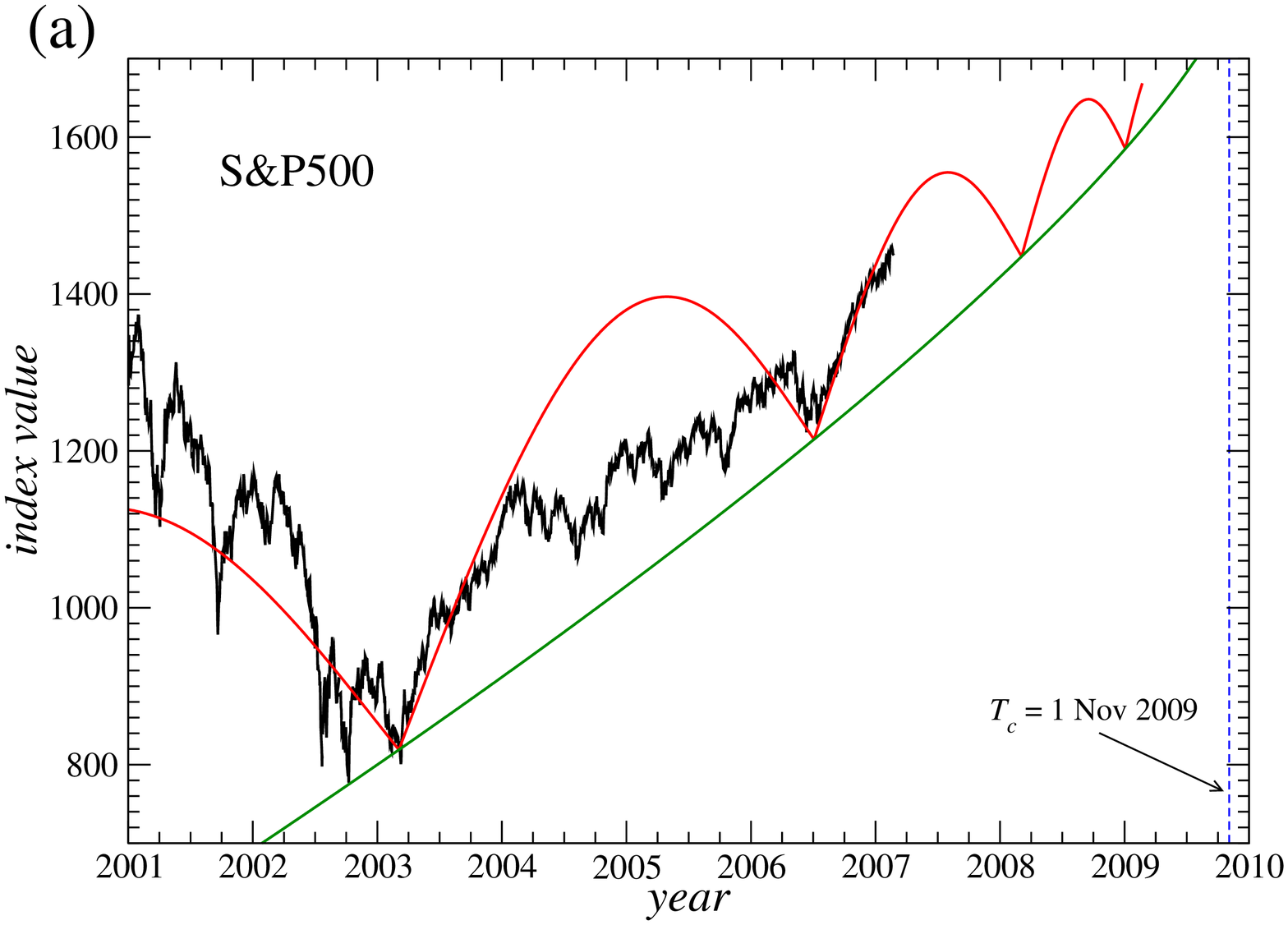}
%\vspace{1cm}
\hspace{-0.5cm}
\includegraphics[width=0.5\textwidth,height=0.5\textwidth]{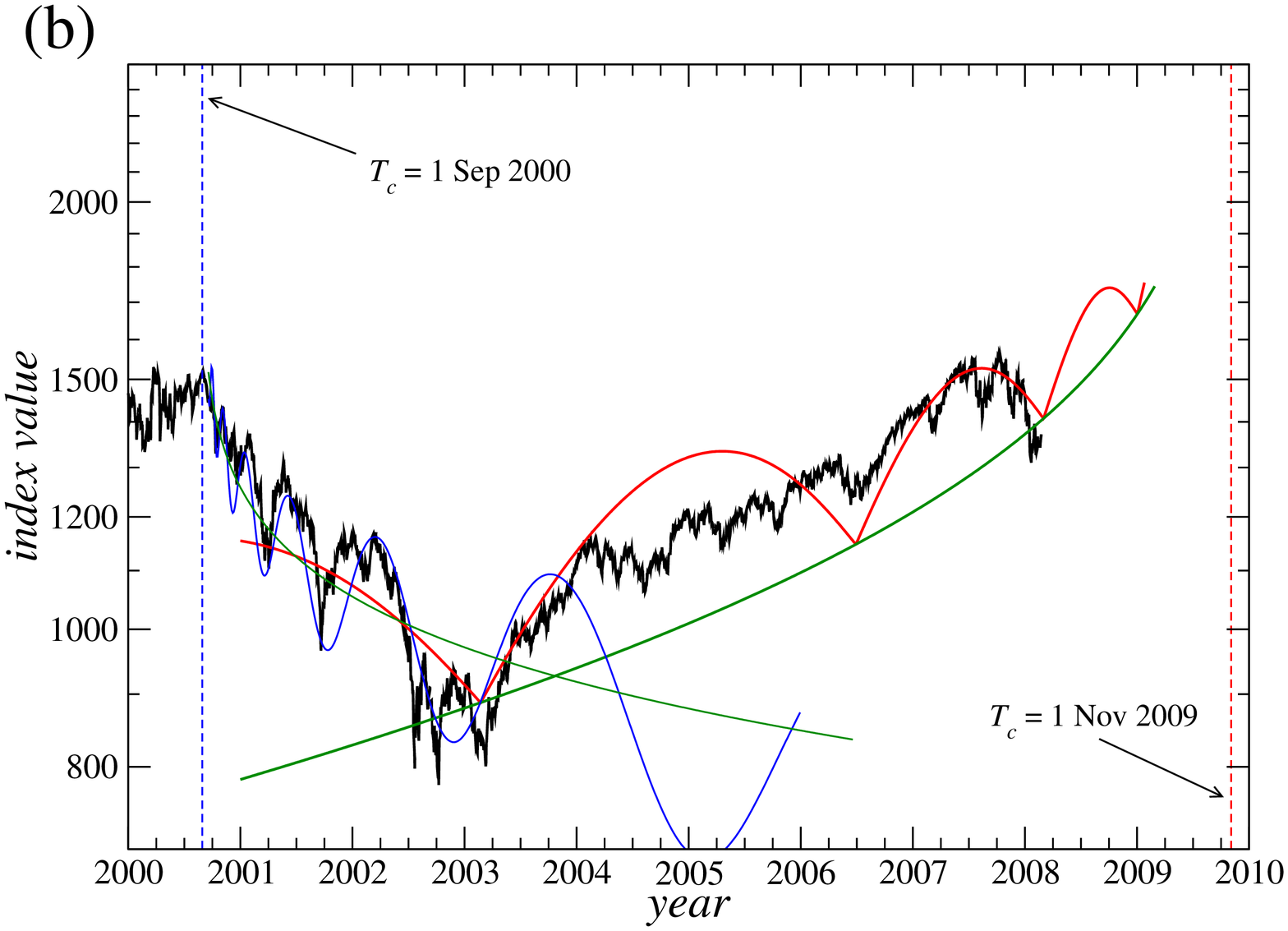}
\caption{(a) Time series of the S$\&$P500 from 1.1.2001 until 23.2.2007 versus the
early (23.2.2007), presented at the FENS2007 Conference, optimal log-periodic representation 
(solid line) with $\lambda=2$ and $T_c$=1.11.2009 for the market bull phase. 
(b) The same log-periodic representation including the S$\&$P500 data up to 23.2.2008.
Only the global trend is corrected as compared to (a). The dashed line indicates the
log-periodically decelerating structure that started on 1.9.2000.}  
\label{fig3}
\end{figure}
%------------------------------------------------------------------------------

An early (made in February 2007, shown also at the time of FENS07 Conference) 
attempt to log-periodically grasp the large scale S$\&$P500 patterns and
to provisionally estimate duration of its present increasing phase, before
it enters recession of comparable magnitude as the one between mid 2000
and early 2003,is shown in Fig.~3. The corresponding critical time $T_c$ points
to the turn of October/November 2009. This scenario thus indicates that until 
this time the market, on average, should preserve its up trend. 
An extra argument in favour of this scenario is that it was predicting 
an intermediate sizeable correction in around the period November 2007- 
February 2008 and it took place indeed. It however also demands that 
in around the end of February 2008 (time of writing this contribution) 
this correction terminates and the index starts rising till at least late
Summer before it starts correcting again. 
The reason for ignoring in this scenario the mid 2004 correction 
seen in the S$\&$P500 and comparable in magnitude to the one (relevant) 
in May-June 2006 is that this former correction is to be interpreted 
a remnant of the bear, since September 2000 log-periodically decelerating 
market component. The related part of this component is also drawn in
the panel (b) of Fig.~3. This panel includes the S$\&$P500 data up to present
and, accordingly, the global trend is updated, but the log-periodic 
structure remains unchanged.
%--------------------------------Figure 4--------------------------------------
\begin{figure}[h]
\begin{center}
%\vspace{1cm} 
%\centerline{\epsfig{figure=fig4.eps,height=8cm,width=10cm}}
%\epsffile{fig1.eps}
\includegraphics[width=0.7\textwidth,height=0.7\textwidth]{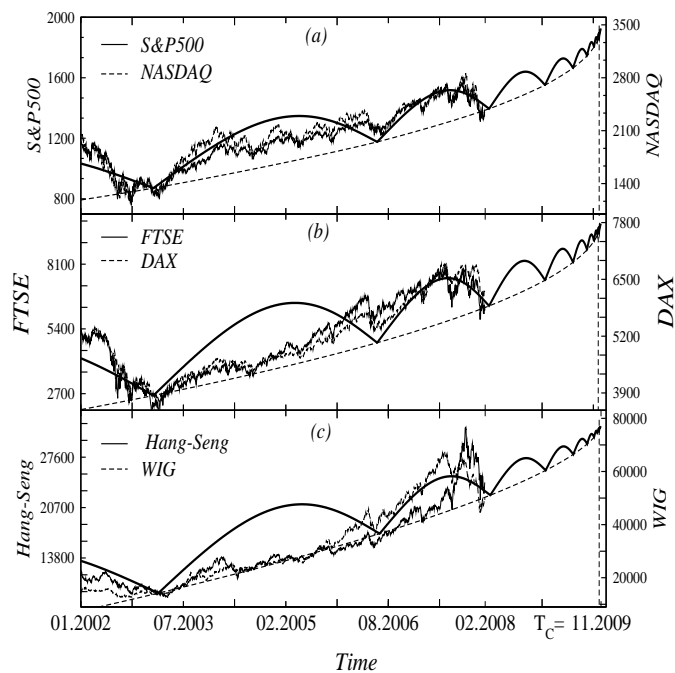}
\caption{(a) Time series from 1.1.2002 until 15.2.2008 for the three pairs of indices:
(a) S$\&$P500 and NASDAQ, (b) DAX and FTSE and (c) HangSeng and WIG, versus the
optimal log-periodic representation for the world bull market phase 
with $\lambda=2$ and $T_c$=1.11.2009.}
\label{fig4}
\end{center}
\end{figure}
%-----------------------------------------------------

The above updated scenario, including the real data up to 
present, for the three pairs of indices, S$\&$P500 and Nasdaq, DAX and FTSE,
HangSeng and WIG (Poland), is shown in the three panels of Fig.~4. 
The log-periodic component remains here the same as the one in Fig.~3.
The main reason for such a selection
of indices presented in Fig.~4 is to show some from among the world most important
ones whose oscillation patterns remain in a satisfactory agreement with the same 
common scenario, as well as those (HangSeng and WIG) whose blind relating to such 
a scenario may seem pointless. Let us recall here however a phenomenon of the
"super-bubble"~\cite{drozdz03}.       
This is an effect that from time to time takes place 
in the financial dynamics and whose identification appears
relevant for a proper interpretation of the financial patterns with
the same universal value of the preferred scaling factor $\lambda=2$. 
This phenomenon of a "super-bubble" is a local boost,   
itself evolving log-periodically, superimposed on top of a long-term bubble
and seen as an extra acceleration of the price increase. Such a "super-bubble"
then crashes and the system returns to a normal bubble state that eventually 
crashes at the time determined by the long-term patterns.   
Two spectacular examples of such a phenomenon are provided by the Nasdaq 
in the first quarter of 2000 and by the gold price
in the beginning of 1981~\cite{drozdz03}. It seems very likely to us that 
the fast increases and then even faster decreases that we see in the last 
two indices of Fig.~4 during the period June 2006 - February 2008 constitute 
further examples of the "super-bubbles" and the corresponding markets have just
returned to their normal long-term (since March 2003) up trend and an ultimate 
critical time $T_c$ may very well coincide with the same common scenario.     
 %-----------------------------------Figure 5-------------------------------------
\begin{figure}
\begin{center}
\includegraphics[width=0.8\textwidth,height=0.7\textwidth]{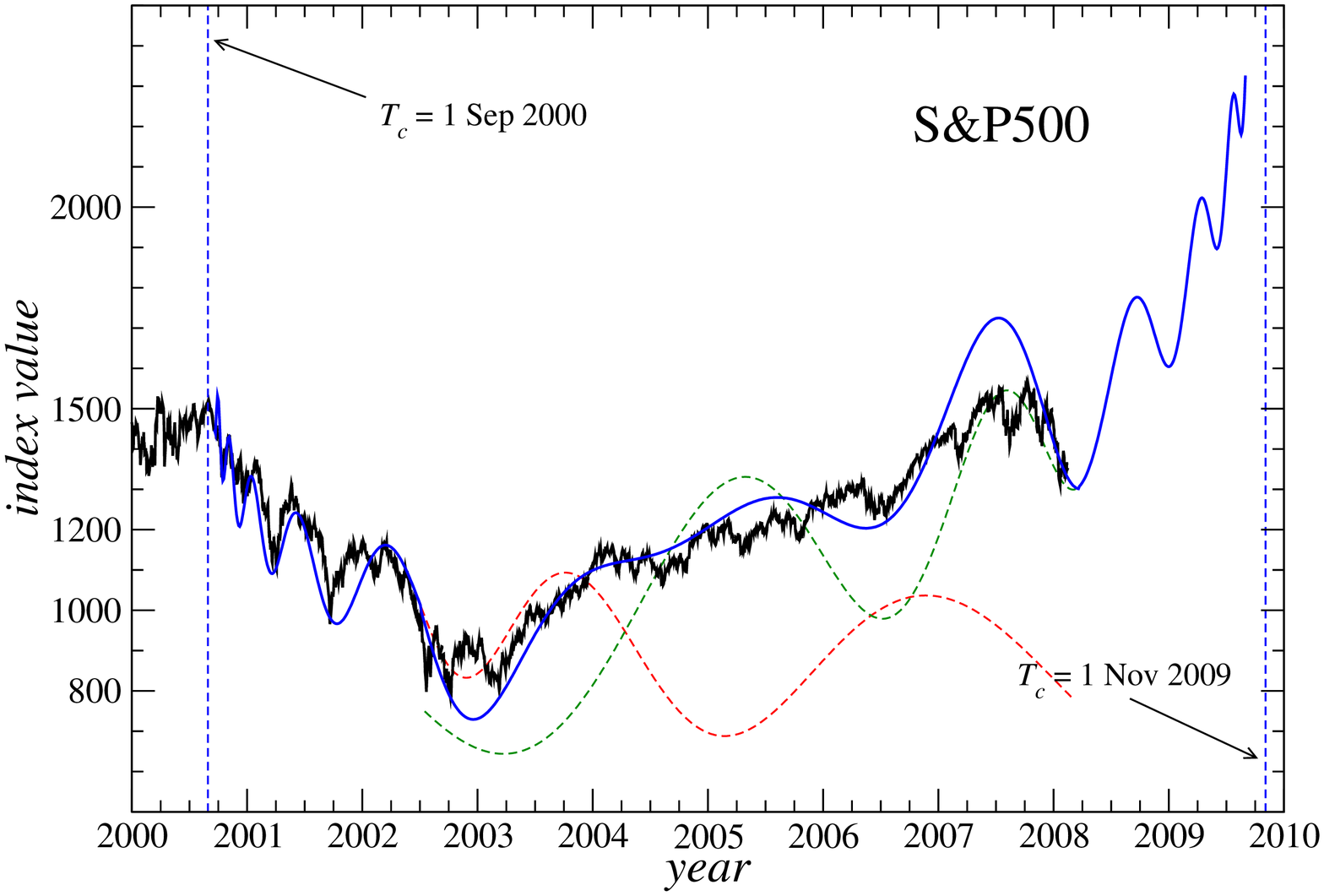}
%\includegraphics[width=6cm, angle=270]{fig6.eps}
%\epsffile{fig1.eps}
\caption{A log-periodic scenario, represented by the solid line,
for the S$\&$P500 development until end 2009. 
The critical time is fixed as $T_c=1.11.2009$, consistently with the 
signal from Fig.~3. 
This solid line is obtained by properly weighting and superimposing
the two $\lambda = 2$ components (dashed lines): 
log-periodically decelerating since 1.9.2000
and log-periodically accelerating towards 1.11.2009, correspondingly.}    
\label{fig5}
\end{center}
\end{figure}
%----------------------------------------------------------------------------------

In ref.~\cite{drozdz06} a speculative scenario for the stock
market (S$\&$P500) development in the time period 2000 - 2010 was invented 
by representing the market index as a superposition of the two components:
one declining and log-periodically decelerating since September 2000 and
the second one rising and thus log-periodically accelerating. Based on the
longest possible time scale (since 1800) log-periodic representation~\cite{drozdz03}
for this index, the critical time for this second component was provisionally              
estimated in around September 2010. Time verified that so far this scenario
makes a lot of sense. In particular, a spectacular increase in 2007 
as well as the reverse of the increasing trend by the end of the year 
was predicted. Recall that this scenario was presented in November 2004.
These facts encourage further its elaboration. 
Since, as discussed above, we now have a more precise estimate for the end
of the present long-term bull market phase the scenario under consideration
can be improved. An accordingly updated variant of this scenario is shown 
in Fig.~5. Interestingly, during the year 2007 this theoretical market 
representation opened room for an even stronger increase than what the S$\&$P500
has performed. Some other markets, like the ones shown in the panel (c)
of Fig.~4, made however use of this freedom and executed a more proportional 
detour.  
%------------------------------------Figure 6-------------------------------------
\begin{figure}
\begin{center}
\includegraphics[width=0.7\textwidth,height=0.5\textwidth]{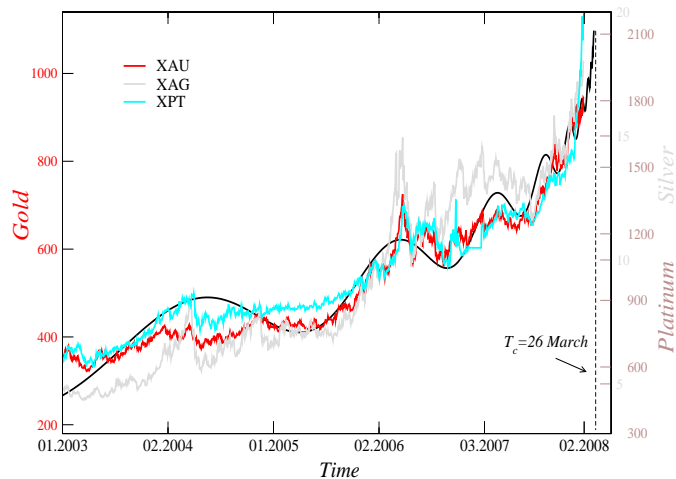}
\caption{Time series of the gold (red), silver (grey) and platinum (cyan) 
prices (in USD) since 1.1.2003 until 22.2.2008, 
versus their optimal log-periodic representation 
with $\lambda = 2$. The corresponding critical time $T_c$=26.3.2008.}    
\label{fig6}
\end{center}
\end{figure}
%----------------------------------------------------------------------------------
\section{Precious metals market}

Some preliminary evidence already exists that the log-periodic oscillations
may also accompany the commodities market dynamics during some more speculative
periods. Such patterns have been identified on the gold market~\cite{drozdz03} 
during the time period 1978-1982, in this case including a very spectacular 
"super-bubble" before the ultimate reverse of the long-term trend, and on the 
oil futures market in the years 1998-2004~\cite{drozdz06}. In this connection
it needs to be also emphasized that the same preferred scaling factor 
$\lambda \approx 2$ as for the stock market turns out here appropriate. 

After many years of stagnation the precious metals market revitalized
starting approximately in the second half of the 2000-2003 stock market
recession period. The following development, at least partly driven by 
speculative activity, elevated the precious metals prices in a relatively short
period of time to the level a factor of 3 to 4 higher. This is illustrated
in Fig.~6 which shows the gold, the silver and the platinum price changes during the period
January 2003 - February 2008. The oscillation patterns of all these prices resemble
each other and, in addition, they quite convincingly coincide with the log-periodic
$(\lambda=2)$ structure which points to the turn of March/April 2008 as the critical 
time $T_c$. Whether this indicates an ultimate reversal of the precious metals
present price trends in around this time or only a sizeable (20-30$\%$) correction
cannot be settled at this stage of the development. If the later possibility 
is to occur, i.e., after such a correction the market starts returning to the
previous highs, 
then - in the spirit of self-similar log-periodicity~\cite{drozdz99} - it is expected
to assume increase towards significantly higher levels and this market phase should
last for another 3-4 years in order to complete the full log-periodic cycle on the
appropriately longer time horizon, such that the whole log-periodic structure seen 
in Fig.~6 constitutes its self-similarly nested substructure. It seems also likely
that the last few months accelerated increases, especially on the platinum market,
can be interpreted in terms of a "super-bubble".   
In any case however at the $T_c$ indicated above one may expect the beginning 
of a significant correction on the precious metals market.   

\section{Oil market}
(Note added on June 23, 2008) Time satisfactorily verified the above
prediction for the
precious metals market. This encourages applying the same methodology to
another
commodity market - the oil market - where the prices expressed in terms of
the USD went
up almost a factor of three over the past one year. Our log-periodic
interpretation of
the underlying price dynamics over the time period 2000-2010 is illustrated
in Fig.~7.
The past year sharp increase turns out to be transparently log-periodic with
$\lambda=2$
and the related critical time corresponds to the first decade of July 2008.
Consequently,
during the Summer that follows the oil prices are likely to drop down even
below 100 USD.
From the longer time perspective, since about 2000, this "super-bubble" decay
is however
going not to constitute an ultimate reverse of the oil up trend. On this
longer time
scale the same universal preferred scaling factor $\lambda=2$ opens room for
the
continuation of increases to similar or even somewhat higher levels and the
estimated
critical time setting the dead-line for this long term increase corresponds
to the late
Summer in 2010, thus several months after the stock market enters a serious
recession.

%------------------------------------Figure 7---------------------------------
\begin{figure}
\begin{center}
\includegraphics[width=0.7\textwidth,height=0.5\textwidth]{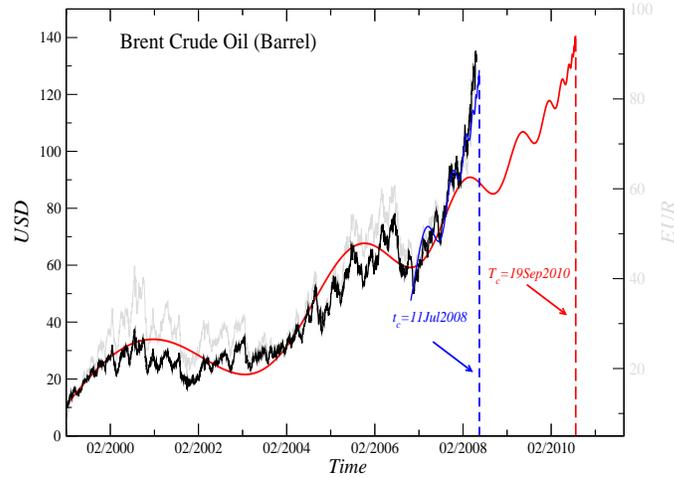}
\caption{Time series of the oil price in USD (grey line represents the oil
price in Euro)
since mid 1999 versus its optimal log-periodic representation with $\lambda =
2$. The
critical time $T_c$=September 2010 corresponds to an ultimate reverse of the
long term
(since 1999) oil price up trend while $t_c$=11.7.2008 sets an upper limit for
the end of
the present "super-bubble".} \label{fig7}
\end{center}
\end{figure}
%-----------------------------------------------------------------------------

\vspace{1cm}

JS thanks the Foundation for Polish Science for financial support through the
Alexander von Humboldt Honorary Research Fellowship.

\end{document}